# Some problems of low-dimensional physics
## (General introduction to low-dimensional quantum movements in many-body systems)

Y. Kornyushin[a)]

*Maître Jean Brunschvig Research Unit, Chalet Shalva, Randogne, CH-3975*

Fermi and kinetic energy are usually calculated in periodic boundary conditions model, which is not self-consistent for low-dimensional problems, when particles are confined. Thus for confined particles the potential box model was used here self-consistently to calculate Fermi and kinetic energies in 3-, 2-, and 1-dimensional cases. This approach is much logical and self-consistent. The criteria of neglecting dimensions, that is the criteria of the 2- and 1-dimensional quantum movements, were derived.

## 1. INTRODUCTION

Fermi and kinetic energies are basic concepts in many-body systems of free identical particles at low temperature, when free identical particles are degenerate. In solid state physics a 3-dimensional case is usually studied.[1] Now 2- and 1- dimensional systems, like thin films and nanowires made of semiconductors, graphenes, carbon nanotubes, nanofibers and others are being studied extensively. These systems are rather promising in many branches of research and technology. Quantum movements in 2- and 1-dimensional systems of free identical particles are different from those in a 3-dimensional one. Fermi and kinetic energies in 2- and 1-dimensional systems of free identical particles are not the same as in a 3-dimensional one. The purpose of this paper is to derive expressions for Fermi and kinetic energies in 2- and 1- dimensional systems of free identical particles, and to obtain criteria of 2- and 1- dimensional quantum movements of free identical particles in a many-body system. These criteria restrict not only dimensions of a sample, but also restrict temperature range.

To perform calculations of Fermi and kinetic energies in a degenerate system of free identical particles the zero boundary conditions model (ZBCM) was formulated and used in this paper. One should be aware that the calculations performed in the present paper are not the exact ones, but they are done in the frame of proposed ZBCM. So obtained results are reliable as far as the ZBCM is relevant.

## 2. FERMI AND KINETIC ENERGIES

To calculate Fermi and Kinetic energies the periodic boundary conditions model (PBCM) is used as a rule.[1] This model is rather adequate for a bulk solid. But when the motion of the particles is confined, the PBCM does not look like a proper one. More adequate in this case is the potential box model. When the potential box is infinitely deep, zero boundary conditions on the borders of the box are realized, thus leading to ZBCM.

2.1. *Fermi and kinetic energies in a 3-dimemsional ZBCM*

In ZBCM an ensemble of free identical particles, confined in an infinitely deep potential box, is considered.[1] Schrödinger equation for the wave function of a particle $\psi(x,y,z)$ is (provided the value of the potential in the box is zero):

$$-(\hbar^2/2m)\Delta\psi(x,y,z) = E\psi(x,y,z), \qquad (1)$$

where $m$ is the mass of a particle and $\hbar$ is the Planck constant divided by $2\pi$.

The wave functions are described by the following equation:[1]

$$\psi(x,y,z) = (8/abc)[\sin(\pi v_x x/a)][\sin(\pi v_y y/b)][\sin(\pi v_z z/c)]. \; v_i = 1, 2, 3, \ldots \qquad (2)$$

The components of the wave vectors are[1] $k_i = \pi v_i/a_i$ (here $a$, $b$ and $c$ are the dimensions of a box and $i = x, y, z$; $a_x = a$, $a_y = b$, and $a_z = c$).

The energy levels are described by the following relation:[1]

$$E(v) = (\pi^2\hbar^2/2m)[(v_x/a)^2 + (v_y/b)^2 + (v_z/c)^2]. \qquad (3)$$

The energy of a ground state ($v_i = 1$) is not zero: $E_0 = (\pi^2\hbar^2/2m)[(1/a^2) + (1/b^2) + (1/c^2)]$, but it is rather small when $a$, $b$, and $c$ are quite large. The energy levels in ZBCM are:

$$E(v) = (\pi^2\hbar^2/2mV^{2/3})v^2, \qquad (4)$$

where $V = abc$ is the volume of a sample, and $v^2 = [(V^{1/3}/a)v_x]^2 + [(V^{1/3}/b)v_y]^2 + [(V^{1/3}/c)v_z]^2$.

When $N = nV$ confined particles fill in the levels (for spin ½ each level is occupied by two particles), they reach in the process of the filling the limiting value of $v$, $v_F$. As $v_i$ are positive numbers only (see Eq. (2)), one has only 1/8 fraction of the incomplete sphere (in the $v_i$ space, excluding planes corresponding to $v_i = 0$) of a radius $v_F$ to be filled in. This yields the total number of the states filled being equal to $(\pi/3)v_F^3 - 0.75\pi v_F^2$ (as $v_i > 0$ only, and $v_i = 0$ does not exist as was pointed out above). They are all occupied by the particles in consideration. For $N$ much larger than unity (which is a regular case) one has

$$(\pi/3)v_F^3 = nV, \text{ or } v_F = (3nV/\pi)^{1/3}. \qquad (5)$$

Eqs. (4,5) yield for the Fermi energy the following expression:

$$E_F = E(v_F) = (\hbar^2/2m)(3\pi^2 n)^{2/3}. \qquad (6)$$

Eq. (6) yields a well-known result, obtained in PBCM.[1]

When one has $\eta$ particles in a unit volume, the Fermi energy of a system is $E_F = (\hbar^2/2m)(3\pi^2\eta)^{2/3}$. If $d\eta$ particles are added, the increase in the energy of a system is $(\hbar^2/2m)(3\pi^2\eta)^{2/3}d\eta$. As the potential energy in a box is assumed to be zero, the kinetic energy of a system is equal to the total energy. The total energy (per unit volume) is the integral of the last expression on $\eta$ from 0 to $n$. So, for the kinetic energy, $T_k$, we have:

$$T_k = 0.6NE_F. \tag{7}$$

Eq. (7) also yields a well-known result obtained in PBCM.[1]

2.2. *Fermi and kinetic energies in 2-dimensional ZBCM*

In 2-dimensional case Eq. (4) should be written as follows:

$$E(v) = (\pi^2\hbar^2/2mS)v^2, \tag{8}$$

where $S = ab$ and $v^2 = [(S^{1/2}/a)v_x]^2 + [(S^{1/2}/b)v_y]^2$.

In 2-dimensional case only ¼ part of the disk (in the $v_i$ space) is relevant because $v_i$ are positive numbers only. Then, instead of $(\pi/3)v_F^3 = nV$, one has $(\pi/2)v_F^2 = N = n_sS = nV$ ($n_s$ is the number of the confined particles per unit area). Instead of $v_F = (3nV/\pi)^{1/3}$ one has $v_F = (2n_sS/\pi)^{1/2}$. One should also take into account that the third (neglected) dimension of a size $\delta \to 0$ yields (according to Eq. (3)) the following contribution to the energy of a system: $E(\delta) = (\pi\hbar)^2/2m\delta^2$. So the Fermi energy is

$$E_F = E(v_F) + E(\delta) = (\hbar^2/2m)[2\pi n_s + (\pi/\delta)^2]. \tag{9}$$

Eq. (9) was also derived for the case of $N$ much larger than unity. The term $\pi/\delta^2$ in Eq. (9) is a very essential one when $\delta$ is small.

Calculation of the kinetic energy in 2-d case like it was done in Section 1.1 yields

$$T_k = (\hbar^2/2m)[\pi N n_s + (\pi/\delta)^2 N]. \tag{10}$$

The last term (proportional to $N/\delta^2$) in the right-hand part of Eq. (10), is a very essential one when $\delta$ is small.

2.3. *Fermi and kinetic energies in 1-dimensional ZBCM*

In 1-dimensional case Eq. (4) should be written as

$$E(v) = (\pi^2\hbar^2/2ma^2)v^2. \tag{11}$$

In 1-dimensional case only half of the double length (in the $v$ space) is relevant because of the positive $v$ numbers only. Then, instead of $(\pi/3)v_F^3 = nV$ one has $2v_F = N = n_l l$ ($n_l$ is the number of the confined particles per unit length), so $v_F = 0.5n_l l$, and

$$E_F = E(\nu_F) + E(\delta) + E(\delta_{max}) = (\hbar^2/2m)[(\pi n_l/2)^2 + (\pi/\delta)^2 + (\pi/\delta_{max})^2]. \tag{12}$$

Eq. (12) was also derived here for the case of $N$ much larger than unity. Two last terms in the right-hand part of Eq. (12) represent the contribution of the neglected dimensions to the energy of a system ($\delta_{max}$ is the larger size of the neglected dimensions).

The kinetic energy in the 1-dimensional case, calculated as in Section 2.1, is

$$T_k = (\pi^2\hbar^2/2m)N[(n_l^2/12) + (1/\delta^2) + (1/\delta_{max})^2]. \tag{13}$$

The contribution of the neglected dimensions is a very essential one when $\delta$ and/or $\delta_{max}$ are/is small.

2.4. *In a potential box of a finite depth all the energy levels (including the Fermi energy) are lower than those in the box of an infinite depth*

The equations for the wave vectors $k_i$ in a finite depth potential box are as follows[2]

$$k_i a_i + 2\arcsin[\hbar k_i/(2mU)^{1/2}] = \nu_i \pi, \; \nu_i = 1, 2, 3, \ldots, \tag{14}$$

where $U$ is the potential depth.

As Eq. (14) has an additional (comparative to the case of the model of infinitely deep potential box) positive term in the left-hand part of it, one can see that the corresponding $k_i$ and $E(\nu_i) = (\hbar k_i)^2/2m$ are smaller than these in the model of infinitely deep potential box either in PBCM, or in ZBCM. The model of the potential box of a finite depth yields lower (comparative to the case of the model of the infinitely deep potential box) energy levels, and the Fermi energy in this model is also lower.

It is worthwhile to note also that in the model of the finite depth of a potential box the boundary conditions are non-zero. From this follows that the wavelength of each state in this model is larger (comparative to that in the infinitely deep potential box model), and corresponding wave vectors and energies are smaller.

## 3. CRITERIA OF LOW-DINENSIONAL MOVEMENTS

3.1. *A criterion of 2-dimensional quantum movement*

Let one of the dimensions be small: $c = \delta \to 0$. The first excited energy level ($\nu_z = 2$), corresponding to this direction in the ZBCM is $2\pi^2\hbar^2/m\delta^2$ (see Eq. (3)). When this quantity is larger than the 2-d Fermi energy (see Eq. (9)),

$$2\pi^2\hbar^2/m\delta^2 > (\hbar^2/2m)[(2\pi n_s) + (\pi/\delta)^2], \tag{15}$$

the quantum movement could be considered as a 2-d one.

Using Eq. (15) and taking into account that the surface density of the particles $n_s = n\delta$,

one can write

$$\delta < (3\pi/2n_s)^{1/2}, \text{ or } \delta < (3\pi/2n)^{1/3}. \qquad (16)$$

From Eq. (16) follows that the density of the delocalized particles $n$ is the most important parameter in the problem considered. Thus for semiconductor with $n = (3\pi/2)\times10^{15}$ cm$^{-3}$ we have $\delta < 100$ nm. One should say that it is quite a considerable size, much larger than typical sizes of regular nanoobjects. So, quantum movements of delocalized electrons in nano-thin films made of semiconductors are often of a 2-dimensional character. Carbon nanoobjects like fullerene molecules, carbon nanotubes, carbon peapods, and graphenes usually have 4 delocalized electrons per one carbon atom [3]. That is the density of the delocalized electrons in carbon nanoobjects $n$ is about $10^{24}$ cm$^{-3}$, which is two orders of magnitude larger than that in metals, where $n$ is about $10^{22}$ cm$^{-3}$. According to Eq. (16) the size of a carbon nano-sample should be less than 0.1 nm to perform a low-dimensional properties of quantum movement of delocalized electrons. So for regular nanoobjects with sizes larger than 0.1 nm the quantum movements of the delocalized electrons are always essentially 3-dimensional ones.

3.2. *A criterion of 1-dimensional quantum movement*

When two of the three dimensions are small, and the first excited energy level (corresponding to $v = 2$) of the largest of the small dimensions $\delta_{max}$ is $2\pi^2\hbar^2/m\delta_{max}^2$, and it is larger than 1-d Fermi energy (see Eqs. (3,12)),

$$2\pi^2\hbar^2/m\delta_{max}^2 > (\hbar^2/2m)[(\pi n_l/2)^2 + (\pi/\delta)^2 + (\pi/\delta_{max})^2], \qquad (17)$$

quantum movement can be considered as a 1-d one.
From Eq. (17) follows that

$$\delta_{max} < [12/(4 + \delta^2 n_l^2)]^{1/2}\delta. \qquad (18)$$

Taking into account that the linear density of the particles $n_l = n\delta\delta_{max}$, one can see that for the case $\delta_{max} = \delta$, Eq.(18) yields:

$$\delta < 2^{1/2}/n^{1/3}. \qquad (19)$$

For $n = 2^{3/2}\times10^{15}$ cm$^{-3}$ we have $\delta < 100$ nm. This is a very considerable value, much larger than one nanometer. Here also the one-dimensional quantum movement of the delocalized electrons could be realized in nanofibers made of semiconductors and not of metals, like in Section 3.1. Quantum movements of delocalized electrons in carbon nanotubes are always essentially 3-dimensional ones.

3.3. *Temperature criterion of low-dimensional quantum movements*

Low-dimensional movements are realized when the difference between the first excited energy level, corresponding to the small direction in ZBCM, $2\pi^2\hbar^2/m\delta^2$ (see Eq. (3) for the corresponding value of $v = 2$), and the ground state level, $\pi^2\hbar^2/2m\delta^2$ (see Eq. (3) for the corresponding value of $v = 1$), is essentially larger then $kT$ ($k$ here is the Boltzmann constant an $T$ is temperature). From this follows that temperature $T$ should be much lower than $3\pi^2\hbar^2/2km\delta^2$:

$$T << 3\pi^2\hbar^2/2km\delta^2. \qquad (20)$$

For $\delta = 10^{-6}$ cm and $m = 9.1 \times 10^{-28}$ g Eq. (20) yields that $T$ should be much lower than 131 K. This value is not an extreme one.

## 4. DISCUSSION

It is shown here that the ZBCM is rather inherent for the description of free charged particles in a potential box. Obtained results are quite close to those obtained in the PBCM.[1] Criteria of realization of low-dimensional movements are derived, including restriction on the temperature. At a comparatively high temperature the quantum movements of particles can be never considered as low-dimensional ones. At low enough temperatures every concrete object that looks like low-dimensional one should be checked according to simple criteria described in this paper if the movements of the particles in it is really low-dimensional ones.

Quantum movements of delocalized electrons in regular carbon nanoobjects are usually of a 3-dimensional character [3].

It is worthwhile to mention here that some collective movements in carbon nanoobjects are of a classical origin, and they don't have to be low-dimensional ones.[3]

[a] Electronic mail: jacqie@bluewin.ch